\documentclass[10pt,twocolumn,english,aps]{revtex4}
\usepackage[T1]{fontenc}
\usepackage[latin9]{inputenc}
\usepackage{color}
\usepackage{array}
\usepackage{amsmath}
\usepackage{graphicx}
\usepackage{epstopdf}
\usepackage{epsfig}
\usepackage{bm}
\usepackage{epsfig}
\usepackage{subfig}
\usepackage{babel}

\makeatletter \providecommand{\tabularnewline}{\\}
\@ifundefined{definecolor}
 {\usepackage{color}}{}
\def\be{\begin{equation}}
\def\ee{\end{equation}}
\def\bea{\begin{eqnarray}}
\def\eea{\end{eqnarray}}

\@ifundefined{showcaptionsetup}{}{ \PassOptionsToPackage{caption=false}{subfig}}
\makeatother

\begin{document}

\title{Intermodal entanglement in Raman processes}
\author{Biswajit Sen$^{1}$ , Sandip Kumar Giri$^{2}$, Swapan Mandal$^{3}$, C. H.
Raymond Ooi$^{4}$, Anirban Pathak$^{5,6}$}

\affiliation{$^{1}$Department of Physics, Vidyasagar Teachers'
Training College, Midnapore-721101, India\\
$^{2}$Department of Physics, Panskura Banamali College,
Panskura-721152, India\\
$^{3}$Department of Physics, Visva-Bharati, Santiniketan, India\\
$^{4}$Department of Physics, University of Malaya, 50603 Kuala
Lumpur, Malaysia\\
$^{5}$Jaypee Institute of Information Technology, A-10, Sector-62,
Noida, UP-201307, India\\
$^{6}$RCPTM, Joint Laboratory of Optics of Palacky University and
Institute of Physics of Academy of Science of the Czech Republic,
Faculty of Science, Palacky University, 17. listopadu 12, 771 46
Olomouc, Czech Republic}

\begin{abstract}
The operator solution of a completely quantum mechanical Hamiltonian of the
Raman processes is used here to investigate the possibility of obtaining
intermodal entanglement between different modes involved in the Raman
processes (e.g. pump mode, Stokes mode, vibration (phonon) mode and
anti-Stokes mode). Intermodal entanglement is reported between a) pump mode
and anti-Stokes mode, b) pump mode and vibration (phonon) mode c) Stokes
mode and vibration phonon mode, d) Stokes mode and anti-stokes mode in the
stimulated Raman processes for the variation of the phase angle of complex
eigenvalue $\alpha_{1}$ of pump mode $a$. Some incidents of intermodal
entanglement in the spontaneous and the partially spontaneous Raman
processes are also reported. Further it is shown that the specific choice of
coupling constants may produce genuine entanglement among Stokes mode,
anti-Stokes mode and vibration-phonon mode. It is also shown that the two
mode entanglement not identified by Duan's criterion may be identified by
Hillery-Zubairy criteria. It is further shown that intermodal entanglement,
intermodal antibunching and intermodal squeezing are independent phenomena.
\end{abstract}

\maketitle

\section{Introduction}

Entanglement is one of the most important resources for quantum
communication and quantum information processing. For example, it is well
known that entanglement is essential for teleportation, dense coding,
quantum information splitting etc. Thus we need entangled states to perform
various important tasks related to quantum information theory. To do so,
first we need a protocol to check, whether a state generally mixed is
entangled or not? This is a very important issue in quantum information
science and several inseparability criteria have been proposed for this
purpose (\cite{phys rep rev} and references therein). In 1996, Peres \cite
{peres} proposed the first inseparability criterion based on negative
eigenvalues of partial transpose of the composite density operator. This
criterion is sufficient and necessary for the detection of entanglement in
(2x2) and (2x3) dimensional states, but is not necessary for higher
dimensional states (see \cite{duan} and references therein). Since the
pioneering work of Peres, several other inseparability inequalities have
been reported for two mode and multi-mode states \cite{duan}-\cite
{A-Miranowicz-et}. Most of these criteria only provide sufficient condition
of inseparability. Further, these criteria may be classified into two sets
\cite{GSA-Ashoka}: A) set of criteria which cannot be directly tested
through experiments \cite{hungh}-\cite{lee} and B) set of criteria which can
be tested experimentally \cite{peres},\cite{duan},\cite{simon}-\cite
{GSA-Ashoka}. Experimentally testable inequalities involve variance or
higher order moments of some observables. Since the expectation values of
physical observables can be measured experimentally, these set of
inseparability criteria can be tested experimentally.

\begin{figure}[h]
%\begin{centering}
\includegraphics[scale=0.4]{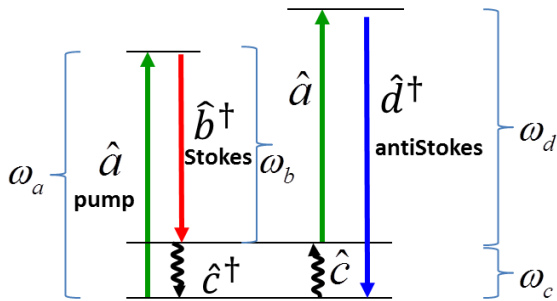} %\par\end{centering}
\caption{(Color online) Two-photon stimulated Raman scheme. The pump photon
is converted into a Stokes photon and a phonon. The pump photon can also mix
with a phonon to produce an anti-Stokes photon.}
\label{fig:scheme}
\end{figure}

The aim of the present work is not to study the inseparability criteria in
detail but to study the possibility of generation of multi-partite entangled
state in two-photon stimulated Raman processes, as depicted in Fig. \ref
{fig:scheme}. The scheme is essentially a sequential double Raman process
that can produce Stokes and antiStokes photons that show highly nonclassical
correlation \cite{nonclassical corr} and macroscopic entanglement in
two-photon laser \cite{two-photon laser}. To study the two-mode entanglement
in the two-photon Raman processes it would be reasonable to use three
criteria from set B. To be precise, we have chosen the two criteria of
Hillery and Zubairy \cite{hz-prl}, \cite{two-mode-citeria-hz} and the
criterion of Duan \emph{et al.} \cite{duan}. Since all these three criteria
are only sufficient, a particular criterion can detect only a subset of all
sets of entangled states. Consequently, application of a single criterion
may yield incomplete result. This is why we have used three experimentally
testable inseparability criteria for our investigation of intermodal
entanglement in stimulated, spontaneous and partially spontaneous Raman
processes.

Nonclassical properties of these Raman processes have been extensively
studied. Initial studies were restricted to the short-time approximation
\cite{szlachetka1}-\cite{perina}. But recently some of the present authors
have reported different nonclassical effects (such as squeezing,
antibunching, intermodal antibunching and sub-shot noise photon number
correlation) in stimulated and spontaneous Raman processes \cite{bsen1}-\cite
{bsen4} without using traditional short-time approximation technique. Our
solution of Raman processes, which does not involve short-time
approximation, is found to reveal many facets of nonclassical effects which
were undetected by short-time approximation technique. However\emph{, }the
possibility of observing intermodal entanglement is not rigorously studied
so far. This fact has motivated us to study the intermodal entanglement in
the double\emph{\ }Raman processes. The present investigation is relevant
for quantum communication for two reasons: Firstly because entanglement is
an essential resource for quantum communication and secondly because
spontaneous Raman process is reported to be useful in the realization of
quantum repeaters \cite{quanreap1}-\cite{quanreap2} which has its
application in long distance high-fidelity quantum communication.

%%%%%%%%%%%%%%%%%%%%%%%%%

%%%%%%%%%%%%%%%%%%%%%%%%
Here it is worthy to note that V Pe\u{r}inov\'{a}\emph{\ et al.} \cite
{perinova} have recently studied the possibility of observing entanglement
in Raman process using the method of invariant subspace. They have followed
an independent approach and have numerically computed the time dependence of
a measure of entanglement. Earlier S V Kuznetsov \emph{et al}. \cite
{kuznetsov} studied the entanglement in the stimulated Raman process
considering only two modes (Stokes mode and phonon mode) and taking the pump
mode as the classical light source. Naturally Kuznetsov \emph{et al}.'s work
illustrated an incomplete scenario and failed to observe intermodal
entanglement involving anti-Stokes mode and pump mode. To circumvent this
limitation we have used here a completely quantum mechanical Hamiltonian.
Further, Pathak, K$\breve{\mathrm{r}}$epelka and Pe\u{r}ina \cite{Anirban
with Perina} have recently investigated the possibilities of observing
intermodal entanglement in the Raman processes using the same Hamiltonian
but with a short-time approximated solution. Their work is restricted by the
intrinsic limitations of the short-time approximation. Such limitations may
be circumvented by the analytical methods developed by us in recent past to
study the stimulated Raman scheme \cite{bsen1}-\cite{bsen4}. Those methods
are systematically used here and a relatively complete scenario of
intermodal entanglement in Raman processes is presented here. Interestingly,
we have observed intermodal entanglement between i) pump mode and
anti-Stokes mode, and ii) Stokes mode and anti-stokes mode. These two
intermodal entanglement was not observed in the earlier analytic studies
\cite{kuznetsov,Anirban with Perina}. The beauty of the present study lies
in the fact that analytic expressions for separability criterion are
obtained by a completely quantum mechanical treatment where all four modes
are considered quantum mechanical. If we look closely into the methodology
adopted in the earlier studies we would quickly find that the approach
adopted in the present paper is simpler and easily extensible to the other
physical systems which are described by bosonic Hamiltonians.

The paper is organized as follows. In the Section \ref{sec:Model-Hamiltonian}
we have described the Hamiltonian of spontaneous and stimulated Raman
processes and its operator solution. In Section \ref
{sec:Intermodal-entanglement} we have used the solution to show that it is
possible to observe intermodal entanglement in Raman processes. The
inseparability criteria used for this purpose are also described in this
section. Finally, Section \ref{sec:Conclusions} is dedicated to conclusions
where we have briefly summarized the result of the present study and have
also discussed the mutual relations among different nonclassical phenomena
observed in the Raman processes.

\section{Model Hamiltonian\label{sec:Model-Hamiltonian}}

The Hamiltonian \cite{szlachetka1}-\cite{bsen4}, \cite{walls} of our
interest is

\begin{equation}
\begin{array}{lcl}
H & = & \omega _{a}a^{\dagger }a+\omega _{b}b^{\dagger }b+\omega
_{c}c^{\dagger }c+\omega _{d}d^{\dagger }d \\
& + & g\left( ab^{\dagger }c^{\dagger }+h.c.\right) +\chi \left(
acd^{\dagger }+\mathrm{h.c.}\right) ,
\end{array}
\label{eq:hamiltonian}
\end{equation}
where h.c. stands for the Hermitian conjugate. Throughout the present paper,
we use $\hbar =1$. The annihilation (creation) operators $a(a^{\dagger
}),\,b(b^{\dagger }),\,c\left( c^{\dagger }\right) ,\,d(d^{\dagger })$
correspond to the laser (pump) mode, Stokes mode, vibration (phonon) mode
and anti-Stokes mode, respectively. They obey the well-known boson
commutation relations. The quantities $\omega _{a}$, $\omega _{b}$, $\omega
_{c}$ and $\omega _{d}$ correspond to the frequencies of pump mode $a$,
Stokes mode $b$, vibration (phonon) mode $c$ and anti-Stokes mode $d$,
respectively. The parameters $g$ and $\chi $ are the Stokes and anti-Stokes
coupling constants, respectively. Coupling constant $g$ ($\chi $) denotes
the strength of coupling between the Stokes (anti-Stokes) mode and the
vibrational (phonon) mode and depends on the actual interaction mechanism.
The dimension of $g$ and $\chi $ are that of frequency and consequently $gt$
and $\chi t$ are dimensionless. Further, $gt$ and $\chi t$ are very small
compared to unity. In order to study the possibility of intermodal
entanglement, we need simultaneous solutions of the following Heisenberg
operator equations of motion for various field operators
\begin{equation}
\begin{array}{lcl}
\dot{a} & = & -i\left( \omega _{a}a+gbc+\chi cd\right) \\
\dot{b} & = & -i\left( \omega _{b}b+gac^{\dagger }\right) \\
\dot{c} & = & -i\left( \omega _{c}c+gab^{\dagger }+\chi a^{\dagger }d\right)
\\
\dot{d} & = & -\left( \omega _{d}d+\chi ac\right) .
\end{array}
\label{eq:equn of motion}
\end{equation}
The above set of coupled nonlinear differential operator equations (\ref
{eq:equn of motion}) are not exactly solvable in the closed analytical form
under weak pump condition. But when the pump is very strong the operator $a$
can be replaced by a $c$-number and the above set of equations (\ref{eq:equn
of motion}) can be solved exactly \cite{perina}. In order to solve these
equations under weak pump approximation we use the perturbative approach.
Our solutions are more general than the well-known short-time approximation.
Details of the calculations are given in our previous papers \cite{bsen1}-
\cite{bsen4}. Here we just note that under weak pump approximation, the
solutions of Eq. (\ref{eq:equn of motion}) assume the following form:
\begin{widetext}
\begin{equation}
\begin{array}{lcl}
a(t) & = & f_{1}a(0)+f_{2}b(0)c(0)+f_{3}c^{\dagger}(0)d(0)+f_{4}a^{\dagger}(0)b(0)d(0)+f_{5}a(0)b(0)b^{\dagger}(0)\\
 & + & f_{6}a(0)c^{\dagger}(0)c(0)+f_{7}a(0)c^{\dagger}(0)c(0)+f_{8}a(0)d^{\dagger}(0)d(0)\\
b(t) & = & g_{1}b(0)+g_{2}a(0)c^{\dagger}(0)+g_{3}a^{2}(0)d^{\dagger}(0)+g_{4}c^{\dagger^{2}}(0)d(0)+g_{5}b(0)c(0)c^{\dagger}(0)\\
 & + & g_{6}b(0)a(0)a^{\dagger}(0)\\
c(t) & = & h_{1}c(0)+h_{2}a(0)b^{\dagger}(0)+h_{3}a^{\dagger}(0)d(0)+h_{4}b^{\dagger}(0)c^{\dagger}(0)d(0)+h_{5}c(0)a(0)a^{\dagger}(0)\\
 & + & h_{6}c(0)b(0)b^{\dagger}(0)+h_{7}c(0)d^{\dagger}(0)d(0)+h_{8}c(0)a^{\dagger}(0)a(0)\\
d(t) & = & l_{1}d(0)+l_{2}a(0)c(0)+l_{3}a^{2}(0)b^{\dagger}(0)+l_{4}b(0)c^{2}(0)+l_{5}c^{\dagger}(0)c(0)d(0)\\
 & + & l_{6}a(0)a^{\dagger}(0)d(0)\end{array}.\label{soln1}\end{equation}
\end{widetext}
The functions $f_{i},\,g_{i},\,h_{i}$ and $l_{i}$ are evaluated from the
dynamics under the initial conditions. In order to apply the boundary
condition, we put $t=0$, in the first term of the Eq. (\ref{soln1}). It is
clear that $f_{1}(0)=g_{1}(0)=h_{1}(0)=l_{1}(0)=1$ and $%
f_{i}(0)=g_{i}(0)=h_{i}(0)=l_{i}(0)=0$ (for $i=2,\,3,\,4,\,5,\,6,\,7$ and $8$%
). Under these initial conditions the corresponding solutions for $%
f_{i}(t),\,g_{i}(t),\,h_{i}(t)$ and $l_{i}(t)$ are obtained as given in the
Appendix.

The solutions Eqs. (\ref{soln1}), (\ref{f})-(\ref{l}) are valid up
to the second orders in $g$ and $\chi $. Interestingly, there is
no restriction on time $t$. For example, $f_{2}$ rises
indefinitely with the increase of time $t.$
Clearly, the divergent nature of the parameters $f_{i},\,g_{i},\,h_{i}$ and $%
l_{i}$ become more explicit as the order of the perturbation theory is
increased. The secular nature is a direct outcome of the perturbation
theory. In the present investigation the secular term is not a problem since
we consider small interaction time. Small interaction time also ensures that
the damping term contributes insignificantly. Here $\Delta \omega
_{1}=\omega _{b}+\omega _{c}-\omega _{a}$ and $\Delta \omega _{2}=\omega
_{a}+\omega _{c}-\omega _{d}$. Normally, the detunings $\Delta \omega _{1}$
and $\Delta \omega _{2}$ are extremely small. In the present investigation,
we however assume that the small (non-zero) detuning is present and hence $%
\Delta \omega _{1}\neq 0$ and $\Delta \omega _{2}\neq 0.$ Here we have used $%
|\Delta \omega _{1}|=0.1$ MHz and $|\Delta \omega _{2}|=0.19$ MHz.
Of course, in Eqs. (\ref{f})-(\ref{l}) we have neglected the terms
beyond the second order in $g$ and $\chi .$ Now we may use Eq.
(\ref{soln1}) to obtain the temporal evolution of the number
operators of various modes as
\begin{widetext}
 \begin{equation}
\begin{array}{lcl}
N_{a}(t) & = & |f_{1}|^{2}a^{\dagger}(0)a(0)+|f_{2}|^{2}b^{\dagger}(0)c^{\dagger}(0)b(0)c(0)+|f_{3}|^{2}c(0)d(0)^{\dagger}c^{\dagger}(0)d(0)\\
 & + & \left[f_{1}^{\ast}f_{2}a^{\dagger}(0)b(0)c(0)+\right.f_{1}^{\ast}f_{3}a^{\dagger}(0)c^{\dagger}(0)d(0)+f_{1}^{\ast}f_{4}a^{\dagger}(0)a^{\dagger}(0)b(0)d(0)\\
 & + & f_{1}^{\ast}f_{5}(a^{\dagger}(0)a(0)+a^{\dagger}(0)a(0)b^{\dagger}(0)b(0))+f_{1}^{\ast}f_{6}a^{\dagger}(0)a(0)c^{\dagger}(0)c(0)\\
 & + & f_{1}^{\ast}f_{7}a^{\dagger}(0)a(0)c^{\dagger}(0)c(0)+f_{1}^{\ast}f_{8}a^{\dagger}(0)a(0)d^{\dagger}(0)d(0)\\
 & + & \left.f_{2}^{\ast}f_{3}b^{\dagger}(0)c^{\dagger^{2}}(0)d(0)+\mathrm{h.c.}\right],\end{array}\label{eq:2.17}\end{equation}

\begin{equation}
\begin{array}{lcl}
N_{b}(t) & = & |g_{1}|^{2}b^{\dagger}(0)b(0)+|g_{2}|^{2}a^{\dagger}(0)c(0)a(0)c^{\dagger}(0)+\left[g_{1}^{*}g_{2}b^{\dagger}(0)a(0)c^{\dagger}(0)\right.\\
 & + & g_{1}^{*}g_{3}b^{\dagger}(0)a^{2}(0)d^{\dagger}(0)+g_{1}^{*}g_{4}b^{\dagger}(0)c^{\dagger^{2}}(0)d(0)+g_{1}^{*}g_{5}(b^{\dagger}(0)b(0)\\
 & + & b^{\dagger}(0)b(0)c^{\dagger}(0)c(0))+g_{1}^{*}g_{6}(b^{\dagger}(0)b(0)+b^{\dagger}(0)b(0)a^{\dagger}(0)a(0))\\
 & + & \left.\mathrm{h.c.}\right],\end{array}\label{2.18}\end{equation}

\begin{equation}
\begin{array}{ccl}
N_{c}(t) & = & |h_{1}|^{2}c^{\dagger}(0)c(0)+|h_{2}|^{2}a^{\dagger}(0)b(0)a(0)b^{\dagger}(0)+|h_{3}|^{2}a(0)d^{\dagger}(0)a^{\dagger}(0)d(0)\\
 & + & \left[h_{1}^{*}h_{2}c^{\dagger}(0)a(0)b^{\dagger}(0)+h_{1}^{*}h_{3}c^{\dagger}(0)a^{\dagger}(0)d(0)\right.\\
 & + & h_{1}^{*}h_{4}c^{\dagger}(0)b^{\dagger}(0)c^{\dagger}(0)d(0)+h_{1}^{*}h_{5}(c^{\dagger}(0)c(0)+c^{\dagger}(0)c(0)a^{\dagger}(0)a(0))\\
 & + & h_{1}^{*}h_{6}(c^{\dagger}(0)c(0)+c^{\dagger}(0)c(0)b^{\dagger}(0)b(0))+h_{1}^{*}h_{7}c^{\dagger}(0)c(0)d^{\dagger}(0)d(0)\\
 & + & \left.h_{1}^{*}h_{8}c^{\dagger}(0)c(0)a^{\dagger}(0)a(0)+h_{2}^{*}h_{3}a^{\dagger^{2}}(0)b(0)d(0)+\mathrm{h.c.}\right],\end{array}\label{2.19}\end{equation}
and

\begin{equation}
\begin{array}{lcl}
N_{d}(t) & = & |l_{1}|^{2}d^{\dagger}(0)d(0)+|l_{2}|^{2}a^{\dagger}(0)a(0)c^{\dagger}(0)c(0)+\left[l_{1}^{\ast}l_{2}a(0)c(0)d^{\dagger}(0)\right.\\
 & + & l_{1}^{\ast}l_{3}a^{2}(0)b^{\dagger}(0)d^{\dagger}(0)+l_{1}^{\ast}l_{4}b(0)c^{2}(0)d^{\dagger}(0)+l_{1}^{\ast}l_{5}c^{\dagger}(0)c(0)d^{\dagger}(0)d(0)\\
 & + & \left.l_{1}^{\ast}l_{6}d^{\dagger}(0)d(0)(1+a^{\dagger}(0)a(0))+\mathrm{h.c.}\right].\end{array}.\label{2.20}\end{equation}
\end{widetext}
In the following section these number operators will be used to study the
intermodal entanglement in Raman processes.

\section{Intermodal entanglement\label{sec:Intermodal-entanglement}}

In order to investigate the intermodal entanglement for various coupled
modes, we assume that all photon and phonon modes are initially coherent. In
other words, the composite boson field consisting of photons and phonon are
in the initial coherent state. Therefore, the composite coherent state
arises from the product of the coherent states $|\alpha _{1}\rangle
,\,|\alpha _{2}\rangle $, $|\alpha _{3}\rangle ,$ and $|\alpha _{4}\rangle $
which are the eigenkets of $a,\,b,\,c$ and $d$ respectively. Thus the
initial composite state is
\begin{equation}
|\psi (0)\rangle =|\alpha _{1}\rangle \otimes |\alpha _{2}\rangle \otimes
|\alpha _{3}\rangle \otimes |\alpha _{4}\rangle .  \label{eq:initial state}
\end{equation}
It is clear that the initial state is separable. Now the field operator $%
a(t) $ operating on such a multi-mode coherent state gives rise to the
complex eigenvalue $\alpha _{1}(t).$ Hence we have,
\begin{equation}
a(0)|\psi (0)\rangle =\alpha _{1}|\psi (0)\rangle ,  \label{3.7}
\end{equation}
where $|\alpha _{1}|^{2}$ is the number of input photons in the pump mode $%
a. $ In a similar fashion we have three more complex amplitudes $\alpha
_{2}(t)$, $\alpha _{3}(t)$ and $\alpha _{4}(t)$ corresponding to the Stokes,
vibrational (phonon) and anti-Stokes field mode operators $b,$ $c$ and $d$
respectively. Clearly, for a spontaneous process, the complex amplitudes are
$\alpha _{2}=\alpha _{3}=\alpha _{4}=0$ and $\alpha _{1}\neq 0.$ For partial
spontaneous process, the complex amplitude $\alpha _{1}$ and any one of the
remaining three eigenvalues are not equal to zero while the other two
complex amplitudes are zero. On the other hand, for a stimulated process,
the complex amplitudes are not necessarily zero. In our present
investigation we consider $\alpha _{1}=\left| \alpha _{1}\right| e^{-i\phi }$
and the other eigenvalues for the Stokes, vibrational (phonon) and
anti-Stokes field modes are real. The aim of the present work is to
investigate the possibility of intermodal entanglement in the spontaneous,
partially spontaneous and stimulated Raman processes. To do so let us begin
with the investigation of two mode entanglement using Hillery and Zubairy's
criteria.

\subsection{Two mode entanglement \label{sub:Two-mode-entanglement}}

There are two criteria due to Hillery and Zubairy \cite{hz-prl}-\cite
{two-mode-citeria-hz}. The first one is
\begin{equation}
\langle N_{a}N_{b}\rangle -|\langle ab^{\dagger }\rangle |^{2}<0.
\label{eq:HZ-1}
\end{equation}
On the other hand, the second criterion is given by
\begin{equation}
\langle N_{a}\rangle \langle N_{b}\rangle -|\langle ab\rangle |^{2}<0.
\label{eq:HZ-2}
\end{equation}
From here onward we will refer to these criteria as HZ-1 and HZ-2 criterion
respectively. In addition to these two criteria, we will also use the Duan's
inseparability criterion due to Duan \emph{et al.} \cite{duan} :

\begin{equation}
\langle \Delta a^{\dagger }\Delta a\rangle \langle \Delta b^{\dagger }\Delta
b\rangle -|\langle \Delta a\Delta b\rangle |^{2}<0.  \label{eq:duan}
\end{equation}
In the criteria Eqs. (\ref{eq:HZ-1})-(\ref{eq:duan}), $a$ and $b$ are
annihilation operators for two arbitrary modes. They are not limited to the
pump mode and the Stokes mode only.

We note that all the above criteria are only sufficient (not necessary) for
detection of entanglement. Keeping this fact in mind, we have applied all
these criteria to study intermodal entanglement between different modes of
Raman Hamiltonian and have observed intermodal entanglement in various
situations.

\begin{figure*}[h]
\centerline{\includegraphics[width=1.8\columnwidth,draft=false]{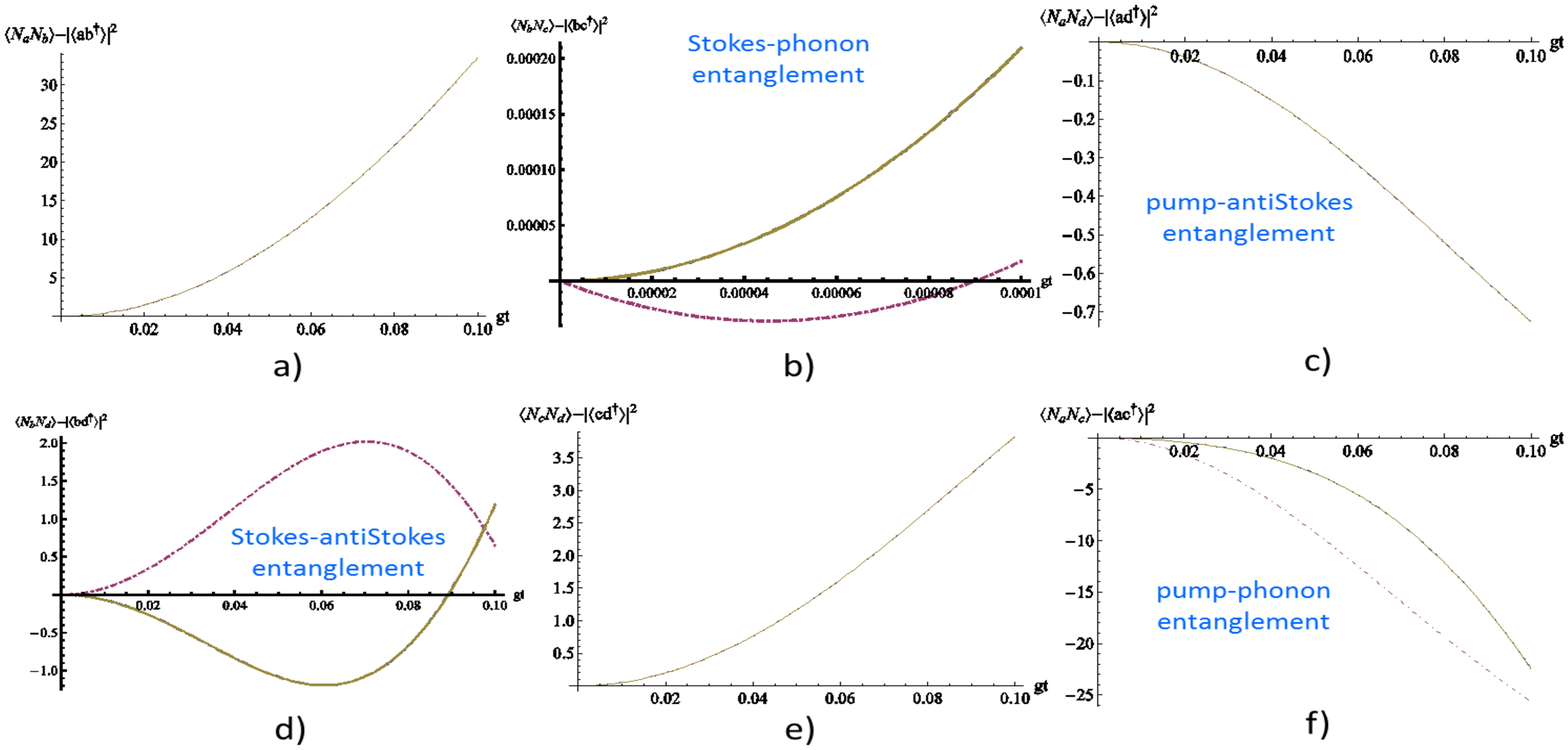}}
\caption{(Color online) Intermodal entanglement in stimulated Raman process
by using HZ-1 criterion:\textcolor{red}{{} }The dotted line, dash-dotted
line and the smooth line are used for the phase angle of the input complex
amplitude $\protect\alpha_{1}$ for $\protect\phi=0,$ $\protect\pi$and $%
\protect\pi/2$ respectively. a) intermodal entanglement is observed between
Stokes mode and vibration-phonon mode for $\protect\phi=\protect\pi/2$, b)
intermodal entanglement is observed between pump mode and anti-Stokes mode,
c) intermodal entanglement is observed between Stokes mode and anti-Stokes
mode for $\protect\phi=0$ and $\protect\pi/2$, d) signature of intermodal
entanglement is not observed between vibration-phonon mode and anti-Stokes
mode, e) intermodal entanglement is observed between pump mode and
vibration-phonon mode.}
\label{fig:HZ1}
\end{figure*}

Let us first investigate the possibility of two mode entanglement in Raman
process using HZ-1 criterion. From Eqs. (\ref{soln1}), (\ref{eq:2.17}), (\ref
{2.18}) and (\ref{eq:initial state}) we obtain

\begin{equation}
\begin{array}{lcl}
\left\langle N_{a}N_{b}\right\rangle -\left| \left\langle ab^{\dagger
}\right\rangle \right| ^{2} & = & \left| f_{3}\right| ^{2}\left| \alpha
_{2}\right| ^{2}\left| \alpha _{4}\right| ^{2} \\
& + & \left| g_{2}\right| ^{2}\left( \left| \alpha _{1}\right| ^{4}-\left|
\alpha _{1}\right| ^{2}\left| \alpha _{2}\right| ^{2}\right) .
\end{array}
\label{abstimulated}
\end{equation}
Consequently, for spontaneous Raman process Eq. (\ref{abstimulated}) reduces
to
\begin{equation}
\begin{array}{lcl}
\left\langle N_{a}N_{b}\right\rangle -\left| \left\langle ab^{\dagger
}\right\rangle \right| ^{2} & = & \left| g_{2}\right| ^{2}\left| \alpha
_{1}\right| ^{4}
\end{array}
\label{eq:Hz1ab}
\end{equation}
It is evidently clear that the right hand side of the Eq. (\ref{eq:Hz1ab})
is always positive. Hence HZ-1 criterion does not show any signature of
intermodal entanglement between pump and Stokes modes in the spontaneous
Raman process. For partially spontaneous Raman process ($|\alpha _{1}|\neq
0,\,|\alpha _{2}|\neq 0,\,|\alpha _{3}|=|\alpha _{4}|=0$), the Eq. (\ref
{abstimulated}) reduces to
\begin{equation}
\left\langle N_{a}N_{b}\right\rangle -\left| \left\langle ab^{\dagger
}\right\rangle \right| ^{2}=\left| g_{2}\right| ^{2}\left( \left| \alpha
_{1}\right| ^{2}-\left| \alpha _{2}\right| ^{2}\right) \left| \alpha
_{1}\right| ^{2}.  \label{eq:partialHZ1ab}
\end{equation}
It is clear that the entanglement is possible in the partially spontaneous
Raman process only when $\left| \alpha _{2}\right| ^{2}>\left| \alpha
_{1}\right| ^{2}$, i.e. the number of Stokes photon is more than the number
of pump photons, which is not the usual case. According to the HZ-1
criterion of Eq. (\ref{eq:HZ-1}), it is clear that the negative values on
the right hand side of Eq. (\ref{abstimulated}) would indicate the presence
of intermodal entanglement between the pump mode and the Stokes mode in
stimulated Raman process. To investigate the possibility of intermodal
entanglement in the stimulated Raman process we have used $\chi =g=10^{5}$
Hz, $|\alpha _{1}|=10,$ $|\alpha _{2}|=8,$ $|\alpha _{3}|=0.01$ and $|\alpha
_{4}|=1$ \cite{foot2}. We have plotted the right hand side of (\ref
{abstimulated}) in Fig. \ref{fig:HZ1}a which does not show any signature of
intermodal entanglement between pump mode and the Stokes mode in the
stimulated Raman process.

%%%%%%%%%%%%%%%%%%%%%%%%%%%%%%%%%%%%%%%%%%%%%%%%%%%
%%%%%%%%%%%%%%%%%%%%%%%%%%%%%%%%%%%%%%%%%%%%%%%%%%%
%%%%%%%%%%%%%%%%%%%%%%%%%%%%%%%%%%%%%%%%%%%%%%%%%%%
\begin{figure*}[h]
\centerline{\includegraphics[width=1.8\columnwidth,draft=false]{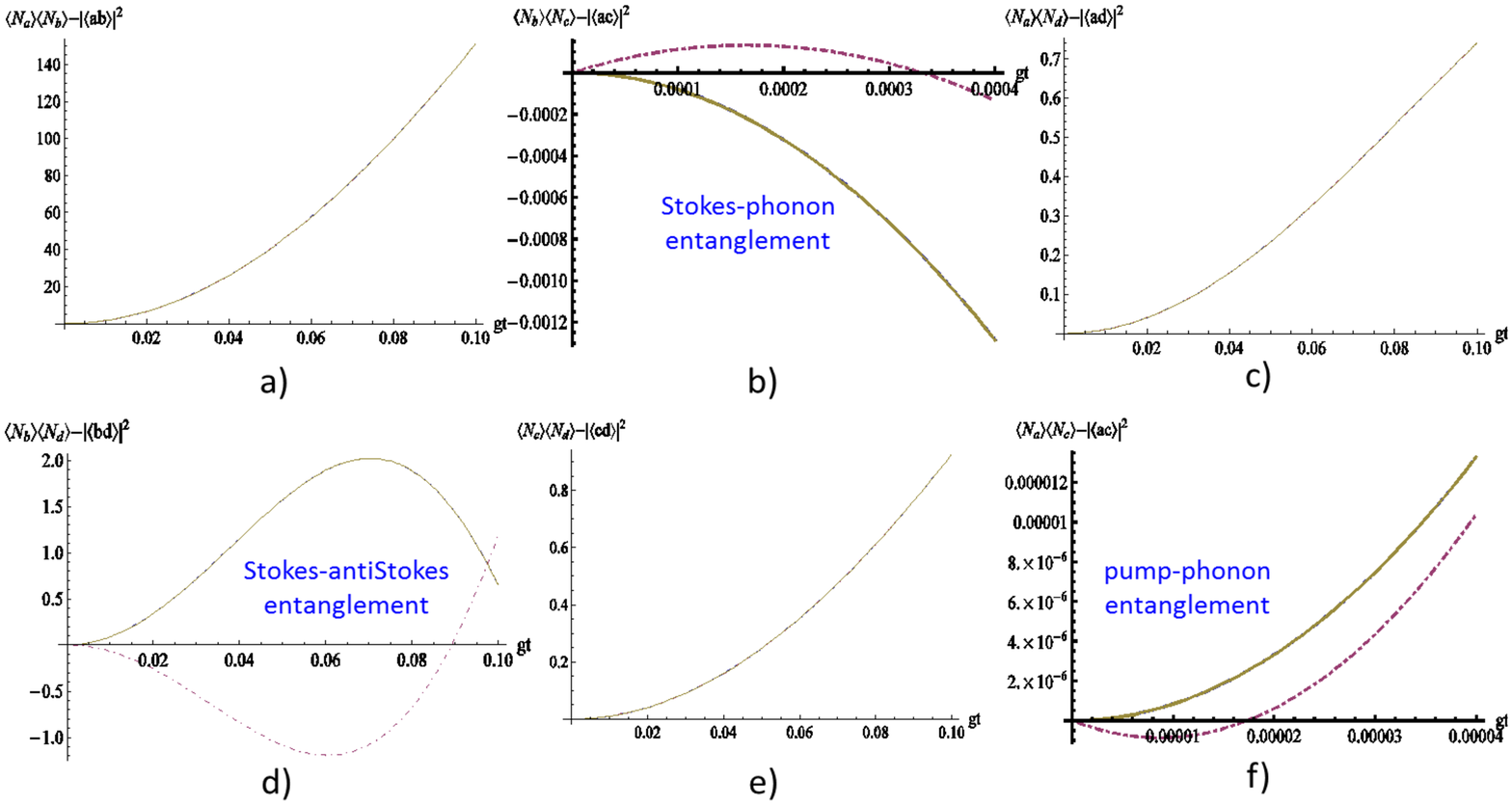}}
\caption{(Color online) Intermodal entanglement in stimulated Raman process
by using HZ-2 criterion: The dotted line, dash-dotted line and the smooth
line are used for the phase angle of the input complex amplitude $\protect%
\alpha_{1}$ for $\protect\phi=0,$ $\protect\pi/2$ and $\protect\pi$
respectively. a) intermodal entanglement is not observed between Stokes mode
and pump mode b) intermodal entanglement is observed between Stokes mode and
vibration-phonon mode, c) signature of intermodal entanglement is not
observed between pump mode and anti-Stokes mode, d) intermodal entanglement
is observed between Stokes mode and anti-Stokes mode only for $\protect\phi=%
\protect\pi/2$, e) intermodal entanglement is not observed between
vibration-phonon mode and anti-Stokes mode, f) intermodal entanglement (time
dependent) is observed between pump mode and vibration-phonon mode for $%
\protect\phi=\protect\pi/2$.}
\label{fig:HZ2}
\end{figure*}
%%%%%%%%%%%%%%%%%%%%%%%%%%%%%%%%%%%%%%%%%%%%%%%%%%%

Here we would like to note that once we have an analytic expression for the
HZ-1 or HZ-2 or Duan criteria in stimulated Raman process, it is
straightforward to study the special cases of: i) spontaneous Raman process,
where $\alpha _{2}=\alpha _{3}=\alpha _{4}=0$ but $\alpha _{1}\neq 0$ and
ii) partial spontaneous Raman process, where $\alpha _{1}\neq 0$ and any one
of the other three $\alpha _{i}\,(i=2,\,3,\,4)$ is non-zero.

The same technique used in the above case is now adopted to obtain the
following equations for the study of intermodal entanglement in stimulated
Raman process using HZ-1 criterion:
\begin{widetext}
\begin{equation}
\begin{array}{lcl}
\left\langle N_{b}N_{c}\right\rangle -\left|\left\langle bc^{\dagger}\right\rangle \right|^{2} & = & \left|g_{2}\right|^{2}\left(3\left|\alpha_{1}\right|^{2}\left|\alpha_{3}\right|^{2}+3\left|\alpha_{1}\right|^{2}\left|\alpha_{2}\right|^{2}+\left|\alpha_{1}\right|^{2}-\left|\alpha_{2}\right|^{2}\left|\alpha_{3}\right|^{2}\right)+\left|h_{3}\right|^{2}\left|\alpha_{2}\right|^{2}\left|\alpha_{4}\right|^{2}\\
 & + & \left[\left\{ h_{1}^{*}h_{2}\alpha_{1}\alpha_{2}^{*}\alpha_{3}^{*}+2g_{4}^{*}g_{1}\alpha_{2}\alpha_{3}^{2}\alpha_{4}^{*}+h_{2}h_{3}^{*}\alpha_{1}^{2}\alpha_{2}^{*}\alpha_{4}^{*}\right\} +{\rm c.c.}\right]\end{array}\label{bc}\end{equation}

\begin{equation}
\begin{array}{lcl}
\left\langle N_{a}N_{d}\right\rangle -\left|\left\langle ad^{\dagger}\right\rangle \right|^{2} & = & \left|f_{3}\right|^{2}\left(\left|\alpha_{3}\right|^{2}+\left|\alpha_{4}\right|^{4}+\left|\alpha_{1}\right|^{2}\left|\alpha_{3}\right|^{2}-\left|\alpha_{1}\right|^{2}\left|\alpha_{4}\right|^{2}\right)-\left|l_{2}\right|^{2}\left(\left|\alpha_{3}\right|^{2}+\left|\alpha_{1}\right|^{2}\left|\alpha_{3}\right|^{2}\right)\end{array}\label{ad}\end{equation}

\begin{equation}
\begin{array}{lcl}
\left\langle N_{b}N_{d}\right\rangle -\left|\left\langle bd^{\dagger}\right\rangle \right|^{2} & = & \left|g_{2}\right|^{2}\left|\alpha_{1}\right|^{2}\left|\alpha_{4}\right|^{2}+\left[\left\{ l_{1}^{*}l_{3}\alpha_{1}^{2}\alpha_{2}^{*}\alpha_{4}^{*}+{\rm c.c.}\right\} \right]\end{array}\label{bd}\end{equation}

\begin{equation}
\begin{array}{lcl}
\left\langle N_{c}N_{d}\right\rangle -\left|\left\langle cd^{\dagger}\right\rangle \right|^{2} & = & \left|l_{2}\right|^{2}\left(2\left|\alpha_{1}\right|^{2}+2\left|\alpha_{1}\right|^{2}\left|\alpha_{4}\right|^{2}-2\left|\alpha_{4}\right|^{2}-\left|\alpha_{4}\right|^{4}-\left|\alpha_{3}\right|^{2}\left|\alpha_{4}\right|^{2}\right)+\left|h_{2}\right|^{2}\left|\alpha_{1}\right|^{2}\left|\alpha_{4}\right|^{2}\end{array}\label{cd}\end{equation}

\begin{equation}
\begin{array}{lcl}
\left\langle N_{a}N_{c}\right\rangle -\left|\left\langle ac^{\dagger}\right\rangle \right|^{2} & = & \left|f_{2}\right|^{2}\left(2\left|\alpha_{1}\right|^{2}+\left|\alpha_{1}\right|^{4}+\left|\alpha_{1}\right|^{2}\left|\alpha_{3}\right|^{2}-4\left|\alpha_{2}\right|^{2}-2\left|\alpha_{1}\right|^{2}\left|\alpha_{2}\right|^{2}-2\left|\alpha_{2}\right|^{2}\left|\alpha_{3}\right|^{2}\right)\\
 & + & \left|f_{3}\right|^{2}\left(\left|\alpha_{4}\right|^{2}+3\left|\alpha_{3}\right|^{2}\left|\alpha_{4}\right|^{2}+3\left|\alpha_{1}\right|^{2}\left|\alpha_{4}\right|^{2}-\left|\alpha_{1}\right|^{2}\left|\alpha_{3}\right|^{2}\right)+\left[\left\{ f_{1}^{\ast}f_{3}\alpha_{1}^{\ast}\alpha_{3}^{\ast}\alpha_{4}\right.\right.\\
 & + & \left.\left.h_{2}^{\ast}h_{3}\alpha_{1}^{\ast2}\alpha_{2}\alpha_{4}+f_{2}^{\ast}f_{3}\alpha_{2}^{\ast}\alpha_{3}^{\ast2}\alpha_{4}\right\} +{\rm c.c.}\right].\end{array}\label{ac}\end{equation}

\end{widetext}
The right hand sides (RHS) of Eqs. (\ref{bc})-(\ref{ac}) are plotted in Fig.
\ref{fig:HZ1}b - Fig. \ref{fig:HZ1}f. It is interesting to note that the
presence of intermodal entanglement in stimulated Raman process is observed
between i) the Stokes mode and the vibration (phonon) mode (Fig. \ref
{fig:HZ1}b), ii) the pump mode and the anti-Stokes mode (Fig. \ref{fig:HZ1}%
c), iii) the Stokes mode and the anti-Stokes mode (Fig. \ref{fig:HZ1}d) and
iv) the pump mode and vibration mode (Fig. \ref{fig:HZ1}f). However, no
signature of intermodal entanglement is observed in the other two cases
(Figs. \ref{fig:HZ1}a and e). Further, it does not show the presence of
genuine entanglement between any three modes of the system. Interestingly,
with the suitable choice of the complex eigenvalues $\alpha _{i}$ it is
possible to observe the signature of the intermodal entanglement using HZ-1
criteria in partially spontaneous Raman process in several ways but no such
signature is observed for the completely spontaneous Raman process.

Since the HZ-1 criterion is only sufficient, we might have failed to detect
some intermodal entanglement. In an attempt to detect such intermodal
entanglement using HZ-2 criterion (\ref{eq:HZ-2}), we have used Eqs. (\ref
{soln1}), (\ref{eq:2.17})-(\ref{2.20}) and (\ref{eq:initial state}) to
yield:
\begin{widetext}
\begin{equation}
\begin{array}{lcl}
\left\langle N_{a}\right\rangle \left\langle N_{b}\right\rangle -\left|\left\langle ab\right\rangle \right|^{2} & = & \left|g_{2}\right|^{2}\left|\mbox{\ensuremath{\alpha}}_{1}\right|^{4}+\left|f_{3}\right|^{2}\left|\mbox{\ensuremath{\alpha}}_{2}\right|^{2}\left|\mbox{\ensuremath{\alpha}}_{4}\right|^{2}-\left[\left(g_{1}^{\ast}g_{6}\right.\right.\\
 & + & \left.\left.f_{1}^{\ast}f_{2}g_{1}^{\ast}g_{2}\right)\left|\mbox{\ensuremath{\alpha}}_{1}\right|^{2}\left|\mbox{\ensuremath{\alpha}}_{2}\right|^{2}+c.c.\right]\end{array},\label{eq:hz2ab}\end{equation}
 and

\begin{equation}
\begin{array}{lcl}
\left\langle N_{b}\right\rangle \left\langle N_{c}\right\rangle -\left|\left\langle bc\right\rangle \right|^{2} & = & \left|g_{2}\right|^{2}\left|\mbox{\ensuremath{\alpha}}_{1}\right|^{2}\left|\mbox{\ensuremath{\alpha}}_{3}\right|^{2}-\left|h_{2}\right|^{2}\left(1+\left|\mbox{\ensuremath{\alpha}}_{2}\right|^{2}\right)\left|\mbox{\ensuremath{\alpha}}_{1}\right|^{2}+\left|h_{3}\right|^{2}\left|\mbox{\ensuremath{\alpha}}_{2}\right|^{2}\left|\mbox{\ensuremath{\alpha}}_{4}\right|^{2}\\
 & - & \left[\left(h_{1}^{*}h_{2}\alpha_{1}\alpha_{2}^{*}\alpha_{3}^{*}+\left(h_{1}h_{4}^{*}+g_{1}g_{2}^{*}h_{1}h_{3}^{*}\right)\alpha_{2}\alpha_{3}^{2}\alpha_{4}^{*}+h_{2}^{*}h_{3}\alpha_{1}^{*2}\alpha_{2}\alpha_{4}\right.\right.\\
 & + & \left.\left.h_{1}^{*}h_{6}\left|\mbox{\ensuremath{\alpha}}_{2}\right|^{2}\left|\mbox{\ensuremath{\alpha}}_{3}\right|^{2}+g_{1}g_{2}^{*}h_{1}^{*}h_{2}\left|\mbox{\ensuremath{\alpha}}_{1}\right|^{2}\left|\mbox{\ensuremath{\alpha}}_{3}\right|^{2}\right)+c.c.\right],\end{array}\label{eq:hh2bc}\end{equation}

\begin{equation}
\begin{array}{lcl}
\left\langle N_{a}\right\rangle \left\langle N_{d}\right\rangle -\left|\left\langle ad\right\rangle \right|^{2} & = & \left|f_{3}\right|^{2}\left|\alpha_{4}\right|^{4}-\left[l_{1}^{*}l_{6}\left|\mbox{\ensuremath{\alpha}}_{1}\right|^{2}\left|\mbox{\ensuremath{\alpha}}_{4}\right|^{2}+c.c\right],\end{array}\label{eq:hz2ad}\end{equation}

\begin{equation}
\begin{array}{lcl}
\left\langle N_{b}\right\rangle \left\langle N_{d}\right\rangle -\left|\left\langle bd\right\rangle \right|^{2} & = & \left|g_{2}\right|^{2}\left|\alpha_{1}\right|^{2}\left|\alpha_{4}\right|^{2}-\left[l_{1}^{*}l_{3}\alpha_{1}^{2}\alpha_{2}^{*}\alpha_{4}^{*}+c.c\right],\end{array}\label{eq:hz2bd}\end{equation}

\begin{equation}
\begin{array}{lcl}
\left\langle N_{c}\right\rangle \left\langle N_{d}\right\rangle -\left|\left\langle cd\right\rangle \right|^{2} & = & \left|h_{2}\right|^{2}\left|\mbox{\ensuremath{\alpha}}_{1}\right|^{2}\left|\mbox{\ensuremath{\alpha}}_{4}\right|^{2}+\left|h_{3}\right|^{2}\end{array}\left|\mbox{\ensuremath{\alpha}}_{4}\right|^{4}-\left[l_{1}^{*}l_{5}\left|\mbox{\ensuremath{\alpha}}_{3}\right|^{2}\left|\mbox{\ensuremath{\alpha}}_{4}\right|^{2}+c.c\right],\label{eq:hz2cd}\end{equation}

\begin{equation}
\begin{array}{lcl}
\left\langle N_{a}\right\rangle \left\langle N_{c}\right\rangle -\left|\left\langle ac\right\rangle \right|^{2} & = & \left|h_{2}\right|^{2}\left|\mbox{\ensuremath{\alpha}}_{1}\right|^{4}-\left|h_{3}\right|^{2}\left(\left|\mbox{\ensuremath{\alpha}}_{4}\right|^{2}+\left|\mbox{\ensuremath{\alpha}}_{1}\right|^{2}\left|\mbox{\ensuremath{\alpha}}_{4}\right|^{2}\right)+\left|f_{3}\right|^{2}\left|\mbox{\ensuremath{\alpha}}_{3}\right|^{2}\left|\mbox{\ensuremath{\alpha}}_{4}\right|^{2}-\left[\left(h_{1}^{\ast}h_{3}\alpha_{1}^{\ast}\alpha_{3}^{\ast}\alpha_{4}\right.\right.\\
 & + & h_{1}^{\ast}h_{8}\left|\mbox{\ensuremath{\alpha}}_{1}\right|^{2}\left|\mbox{\ensuremath{\alpha}}_{3}\right|^{2}+h_{2}^{\ast}h_{3}\alpha_{1}^{\ast2}\alpha_{2}\alpha_{4}-h_{1}^{\ast}h_{5}\left|\mbox{\ensuremath{\alpha}}_{1}\right|^{2}\left|\mbox{\ensuremath{\alpha}}_{3}\right|^{2}\\
 & + & \left.\left.f_{1}^{\ast}f_{2}h_{3}^{\ast}h_{1}\alpha_{2}\alpha_{3}^{2}\alpha_{4}^{\ast}+f_{1}^{\ast}f_{3}h_{3}^{\ast}h_{1}\left|\mbox{\ensuremath{\alpha}}_{3}\right|^{2}\left|\mbox{\ensuremath{\alpha}}_{4}\right|^{2}\right)+c.c.\right].\end{array}\label{eq:hz2ac}\end{equation}
 \end{widetext}

From the closed form analytic expressions Eqs. (\ref{eq:hz2ab})-(\ref
{eq:hz2ac}), it is possible to obtain the signature of intermodal
entanglement in various cases. Interestingly, we obtain intermodal
entanglement between Stokes mode and vibration mode for spontaneous Raman
process. The intermodal entanglement Eqs. (\ref{eq:hz2ab})-(\ref{eq:hz2ac})
for stimulated Raman processes are illustrated in the Fig. \ref{fig:HZ2}%
a-Fig. \ref{fig:HZ2}f. In accordance to HZ-2 criterion, negative values of
the ordinates indicate the signature of entanglement. Therefore, the
intermodal entanglement is observed in: i) Stokes mode and vibration mode,
ii) Stokes mode and anti-Stokes mode and iii) pump mode and vibration mode.
However, there is no signature of intermodal entanglement in the remaining
cases for Stimulated Raman processes. It is possible to obtain the
intermodal entanglement for various partially spontaneous Raman processes.
However, these results are not exhibited in the present text. It is
interesting to note that HZ-2 criterion failed to detect intermodal
entanglement between, pump and anti-Stokes mode. Thus the Raman process
provides a very nice example of physical system where it can be shown with
physical example that these inseparability criteria are only sufficient.
Still there are two situations where we have not found the signature of
intermodal entanglement. Let us see what happens when we apply another
sufficient but not necessary criterion of inseparability.

%%%%%%%%%%%%%%%%%%%%%%%%%%%%%%%%%%%%%%%%%%%%%%%%%%%
\begin{figure*}[h]
\centerline{\includegraphics[width=1.8\columnwidth,draft=false]{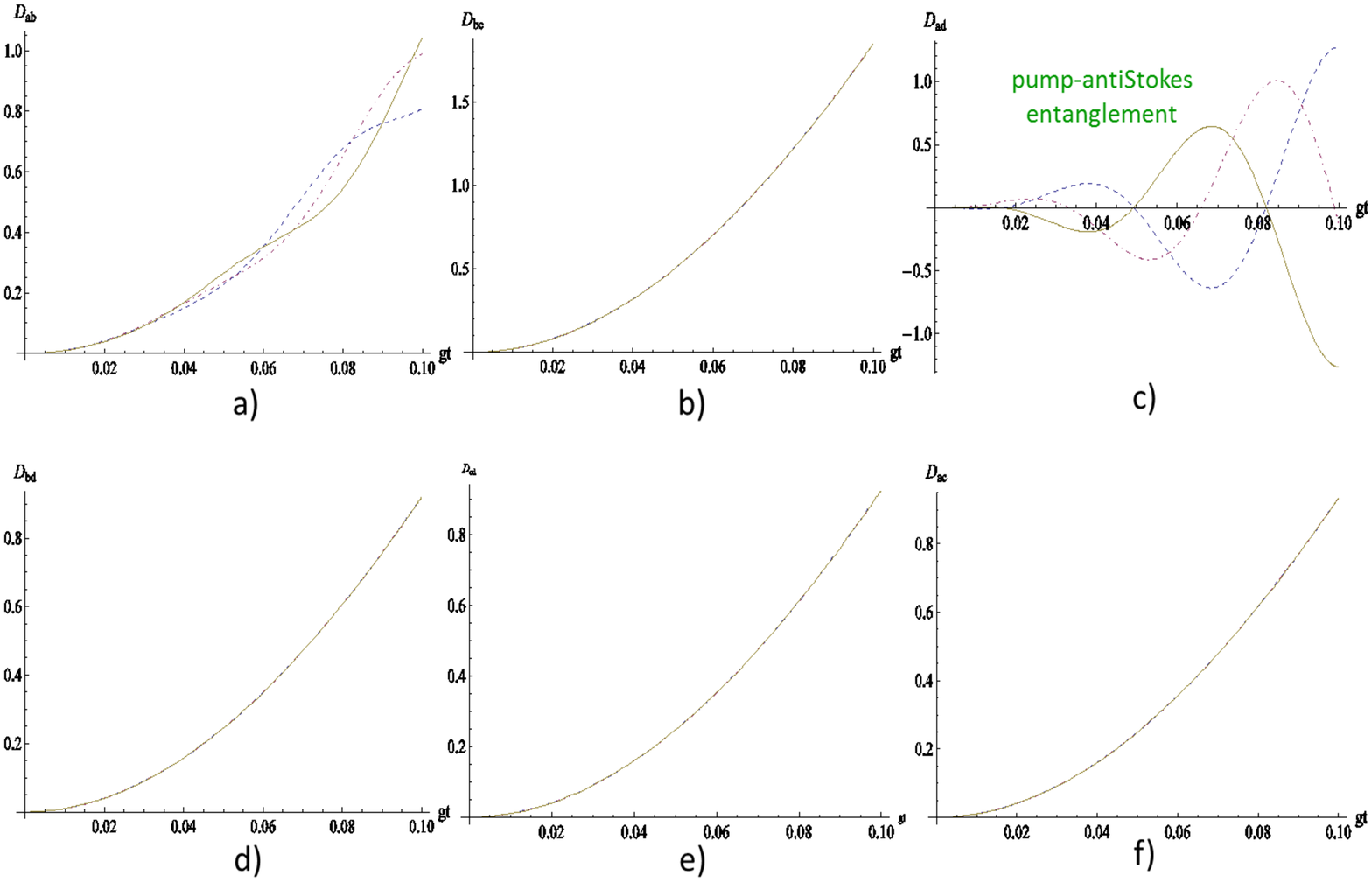}}
\caption{(Color online) Intermodal entanglement in stimulated Raman process
by using Duan criterion:\textcolor{red}{{} }The dotted line, dash-dotted
line and the smooth line are used for the phase angle of the input complex
amplitude $\protect\alpha_{1}$ for $\protect\phi=0,$ $\protect\pi/2$ and $%
\protect\pi$ respectively. Intermodal entanglement is only observed between
pump mode and anti-Stokes mode (Fig.\ref{fig: Duan}.(c)).}
\label{fig: Duan}
\end{figure*}
%%%%%%%%%%%%%%%%%%%%%%%%%%%%%%%%%%%%%%%%%%%%%%%%%%%

Now Duan criterion Eq. (\ref{eq:duan}) for the intermodal entanglement can
also be written as \cite{shivkumar}
\begin{equation}
\begin{array}{lcl}
D=\left\langle \left( \Delta u\right) ^{2}\right\rangle +\left\langle \left(
\Delta v\right) ^{2}\right\rangle & -2< & 0
\end{array}
\label{eq:duan1}
\end{equation}
where
\begin{equation}
\begin{array}{lcl}
\hat{u} & = & \frac{1}{\sqrt{2}}\left[ \left( a+a^{\dagger }\right) +\left(
b+b^{\dagger }\right) \right]
\end{array}
\label{eq:duan2}
\end{equation}
and

\begin{equation}
\begin{array}{lcl}
\hat{v} & = & \frac{1}{i\sqrt{2}}\left[ \left( a-a^{\dagger }\right) +\left(
b-b^{\dagger }\right) \right]
\end{array}
.  \label{eq:duan3}
\end{equation}
Using Eqs. (\ref{soln1}), (\ref{eq:initial state}) and (\ref{eq:duan1})-(\ref
{eq:duan3}) we can obtain analytic expression for $D$ the left hand side of
Duan \emph{et al.} criterion Eq. (\ref{eq:duan1}),

For mode $a$ and $b$,
\begin{equation}
\begin{array}{lcl}
D_{ab} & = & 2\left[\left|f_{3}\right|\left|\alpha_{4}\right|^{2}+%
\left|g_{2}\right|^{2}\left|\alpha_{1}\right|^{2}\right. \\
& + & \frac{1}{2}\left\{
\left(f_{1}g_{6}^{\ast}+f_{5}g_{1}^{\ast}\right)\alpha_{1}\alpha_{2}^{\ast}%
\right. \\
& + & \left.\left.\left(2f_{1}g_{3}^{\ast}+f_{4}g_{1}^{\ast}+f_{3}g_{2}^{%
\ast}\right)\alpha_{1}^{\ast}\alpha_{4}+\mathrm{c.c.}\right\} \right],
\end{array}
\label{eq:d1}
\end{equation}
for mode $a$ and $c$

\begin{equation}
\begin{array}{lcl}
D_{ac} & = & 2\left[\left|f_{3}\right|\left|\alpha_{4}\right|^{2}+%
\left|h_{2}\right|^{2}\left|\alpha_{1}\right|^{2}+\left|h_{3}\right|^{2}%
\left|\alpha_{4}\right|^{2}\right. \\
& + & \frac{1}{2}\left\{
\left(f_{1}h_{5}^{\ast}+f_{6}h_{1}^{\ast}+f_{3}h_{3}^{\ast}\right.\right. \\
& + & \left.\left.\left.f_{7}h_{1}^{\ast}+f_{1}h_{8}^{\ast}\right)\alpha_{1}%
\alpha_{3}^{\ast}\right\} +\mathrm{c.c.}\right],
\end{array}
\label{eq:d2}
\end{equation}
for mode $b$ and $c$
\begin{equation}
\begin{array}{lcl}
D_{bc} & = & 2\left[\left|g_{2}\right|^{2}\left|\alpha_{1}\right|^{2}+%
\left|h_{2}\right|^{2}\left|\alpha_{1}\right|^{2}+\left|h_{3}\right|^{2}%
\left|\alpha_{4}\right|^{2}\right. \\
& + & \left.\frac{1}{2}\left\{
\left(g_{1}h_{6}^{\ast}+g_{5}h_{1}^{\ast}+g_{2}h_{2}^{\ast}\right)%
\alpha_{3}^{\ast}\alpha_{2}\right\} +\mathrm{c.c.}\right],
\end{array}
\label{eq:d3}
\end{equation}
for mode $a$ and $d$
\begin{equation}
\begin{array}{lcl}
D_{ad} & = & 2\left[\left|f_{3}\right|\left|\alpha_{4}\right|^{2}+\frac{1}{2}%
\left\{
\left(f_{1}l_{2}^{\ast}+f_{3}l_{1}^{\ast}\right)\alpha_{3}^{\ast}\right.%
\right. \\
& + & \left.\left.\left(2f_{1}l_{3}^{\ast}+f_{4}l_{1}^{\ast}+f_{2}l_{2}^{%
\ast}\right)\alpha_{1}^{\ast}\alpha_{2}+f_{8}l_{1}^{\ast}\alpha_{1}%
\alpha_{4}^{\ast}+\mathrm{c.c.}\right\} \right],
\end{array}
\label{eq:d4}
\end{equation}
for mode $c$ and $d$
\begin{equation}
\begin{array}{lcl}
D_{cd} & = & 2\left[\left|h_{2}\right|^{2}\left|\alpha_{1}\right|^{2}+%
\left|h_{3}\right|^{2}\left|\alpha_{4}\right|^{2}+\right. \\
& + & \frac{1}{2}\left\{
\left(2l_{4}h_{1}^{\ast}+l_{2}h_{2}^{\ast}+l_{1}h_{4}^{\ast}\right)%
\alpha_{2}\alpha_{3}\right. \\
& + & \left.\left.\left(l_{5}h_{1}^{\ast}+l_{1}h_{7}^{\ast}\right)%
\alpha_{3}^{\ast}\alpha_{4}++\mathrm{c.c.}\right\} \right],
\end{array}
\label{eq:d5}
\end{equation}
and for mode $b$ and $d$

\begin{center}
\begin{equation}
\begin{array}{lcl}
D_{bd} & = & 2\left[\left|g_{2}\right|^{2}\left|\alpha_{1}\right|^{2}+\frac{1%
}{2}\left\{
\left(l_{4}g_{1}^{\ast}+l_{2}g_{2}^{\ast}+l_{1}g_{4}^{\ast}\right)%
\alpha_{3}^{2}+\mathrm{c.c.}\right\} \right]
\end{array}
.  \label{eq:d6}
\end{equation}
\end{center}

Right hand sides of equations (\ref{eq:d1})-(\ref{eq:d6}) are plotted in
Fig. \ref{fig: Duan}a- Fig \ref{fig: Duan}f. It is clear that the
intermodal-entanglement is observed only between the pump mode and
anti-Stokes mode. The criterion is non-conclusive in all other cases. This
is so because Duan criterion is only sufficient.

\begin{widetext}

\begin{table}
\begin{centering}
\begin{tabular}{|>{\centering}p{0.8cm}|>{\centering}p{1.2cm}|>{\centering}p{1.4cm}|>{\centering}p{1.2cm}|>{\centering}p{1.2cm}|>{\centering}p{1.4cm}|>{\centering}p{1.2cm}|>{\centering}p{1.2cm}|>{\centering}p{0.7cm}|c|>{\centering}p{1.6cm}|>{\centering}p{1.6cm}|}
\hline {\scriptsize inter-mode } &
\multicolumn{3}{c|}{{\scriptsize HZ-1}} &
\multicolumn{3}{c|}{{\scriptsize HZ-2}} &
\multicolumn{3}{c|}{{\scriptsize Duan}} & {\scriptsize
antibunching \cite{bsen2}} & {\scriptsize Squeezing\cite{bsen1}
}\tabularnewline \hline
 & {\scriptsize $\phi=0$ } & {\scriptsize $\pi/2$ } & {\scriptsize $\pi$ } & {\scriptsize $\phi=0$ } & {\scriptsize $\pi/2$ } & {\scriptsize $\pi$ } & {\scriptsize $\phi=0$ } & {\scriptsize $\pi/2$ } & {\scriptsize $\pi$ } &  & \tabularnewline
\hline {\scriptsize $ab$ } & {\scriptsize nc } & {\scriptsize nc }
& {\scriptsize nc } & {\scriptsize nc } & {\scriptsize nc } &
{\scriptsize nc } & {\scriptsize nc } & {\scriptsize nc } &
{\scriptsize nc } & {\scriptsize possible } & {\scriptsize
possible}\tabularnewline \hline {\scriptsize $ac$ } & {\scriptsize
entangled } & {\scriptsize entangled } & {\scriptsize entangled }
& {\scriptsize nc } & {\scriptsize entangled (time-dependent) } &
{\scriptsize nc } & {\scriptsize nc } & {\scriptsize nc } &
{\scriptsize nc } & {\scriptsize bunching } & {\scriptsize
possible}\tabularnewline \hline {\scriptsize $ad$ } & {\scriptsize
entangled } & {\scriptsize entangled } & {\scriptsize entangled }
& {\scriptsize nc } & {\scriptsize nc } & {\scriptsize nc } &
{\scriptsize entangled } & {\scriptsize nc } & {\scriptsize nc } &
{\scriptsize anti-bunching } & {\scriptsize
possible}\tabularnewline \hline {\scriptsize $bc$ } & {\scriptsize
nc } & {\scriptsize entangled (time-dependent) } & {\scriptsize nc
} & {\scriptsize entangled } & {\scriptsize nc } & {\scriptsize
entangled } & {\scriptsize nc } & {\scriptsize nc } & {\scriptsize
nc } & {\scriptsize bunching } & {\scriptsize
possible}\tabularnewline \hline {\scriptsize $bd$ } & {\scriptsize
entangled } & {\scriptsize nc } & {\scriptsize entangled } &
{\scriptsize nc } & {\scriptsize entangled } & {\scriptsize nc } &
{\scriptsize nc } & {\scriptsize nc } & {\scriptsize nc } &
{\scriptsize antibunching } & {\scriptsize not
possible}\tabularnewline \hline {\scriptsize $cd$ } & {\scriptsize
nc } & {\scriptsize nc } & {\scriptsize nc } & {\scriptsize nc } &
{\scriptsize nc } & {\scriptsize nc } & {\scriptsize nc } &
{\scriptsize nc } & {\scriptsize nc } & {\scriptsize bunching } &
{\scriptsize possible}\tabularnewline \hline
\end{tabular}
\par\end{centering}

\caption{\label{tab:Table1}Relation between different nonclassical phenomena
observed in stimulated Raman process. Here nc stands for non-conclusive.
{\tiny . }}
\end{table}
\end{widetext}

\section{Conclusions\label{sec:Conclusions}}

We have clearly established that the stimulated Raman process can produce
intermodal entanglement. The observations are summarized in the Table \ref
{tab:Table1}. Here it would be apt to note that recently Pathak, $\mathrm{K}%
\breve{\mathrm{r}}\mathrm{epelka}$ and $\mathrm{Pe\breve{r}ina}$ \cite
{Anirban with Perina} have investigated the possibilities of observing
intermodal entanglement in the Raman processes using a approximated
short-time solution. They have identified intermodal entanglement in \emph{%
pump-phonon} $ac$ and \emph{Stokes-phonon} $bc$ modes only. However, in
addition to those two modes we have observed intermodal entanglement in
\emph{pump-antiStokes }$ad$ and \emph{Stokes-antiStokes} $bd$ modes too. In
addition to these, we explore the various possibilities of getting
intermodal entanglement in partial spontaneous Raman processes too. In this
way, our solution is found more powerful compared to those of the solutions
of Raman processes under short-time approximation. Further, the use of
short-time solution led to the monotonic nature of entanglement parameter as
seen in Eqs. (11) and (12) of ref. \cite{Anirban with Perina}. As our
solution is valid for all times and hence the entanglement parameters are
free from this particular problem which is generally a characteristic of
short-time solutions. Another earlier effort to study the intermodal
entanglement in the Raman processes by S. V. Kuznetsov \cite{kuznetsov} was
restricted to the study of intermodal entanglement between Stokes mode and
the vibration mode as they had considered a simplified two-mode Hamiltonian.
Thus the use of a completely quantum mechanical description of the Raman
process, our solution, and the strategy to use more than one inseparability
criterion have helped us to obtain a relatively more complete picture of the
intermodal entanglement in the Raman processes. Further, if we look at the
possibilities of different kinds of nonclassicalities summarized in Table
\ref{tab:Table1} (see the rows corresponding to $ac$, $bc$, and $bd$ modes),
then we can quickly recognize that the existence of any one of the
nonclassical phenomenon does not depend on the presence of the other. To be
precise, entanglement, antibunching and squeezing are nonclassical phenomena
but they are independent of each other. However, Duan criterion in the
present form implies intermodal squeezing in one of the quadrature variable
but the converse is not true. This fact can be clearly seen in the Table \ref
{tab:Table1}, where we note that except in $bd$ mode intermodal squeezing is
possible in all other coupled modes. However, the Duan criterion of
intermodal entanglement is satisfied only for the $ac$ mode.

In quantum optics, physical systems (matter-field interactions) are usually
described by multi-mode bosonic Hamiltonians. The procedure followed in the
present work may be applied directly to those systems to study the
intermodal entanglement. It is expected that most of these systems will show
intermodal entanglement. This is so because most of the quantum states are
entangled. Separability is a very special case. A natural question should
arise at this point: If entanglement is so common why are we looking for it?
\emph{The answer lies in the fact that entanglement is one of the most
important resources for quantum information processing and quantum
communication but still it is not very easy to produce useful multi-partite
entanglement.} As many of the quantum optical systems described by bosonic
Hamiltonian (including the system studied here) are experimentally
achievable, useful intermodal entanglement may be produced by them.
Entanglement obtained in such a system is expected to find application in
controlled quantum teleportation, quantum information splitting, dense
coding, direct secured quantum communication etc. We conclude this paper
with an optimistic view that the present work will motivate others to look
for theoretical and experimental generation of useful multi-mode
entanglement in other quantum optical systems.

\section{ACKNOWLEDGMENTS}

AP thanks Department of Science and Technology (DST), India for support
provided through the DST project No. SR/S2/LOP-0012/2010 and he also thanks
the Operational Program Education for Competitiveness - European Social Fund
project CZ.1.07/2.3.00/20.0017 of the Ministry of Education, Youth and
Sports of the Czech Republic. RO thanks the Ministry of Higher Education
(MOHE)/University of Malaya HIR Grant No. A-000004-50001 for support.

\appendix
%dummy comment inserted by tex2lyx to ensure that this paragraph is not empty

\section{Parameters for the solutions in Eq. (3)}

\begin{eqnarray}
f_{1} &=&\exp (-i\omega _{a}t),  \notag \\
f_{2} &=&\frac{ge^{-i\omega _{a}t}}{\Delta \omega _{1}}\left[ e^{-i\Delta
\omega _{1}t}-1\right] ,  \notag \\
f_{3} &=&-\frac{\chi e^{-i\omega _{a}t}}{\Delta \omega _{2}}\left[
e^{i\Delta \omega _{2}t}-1\right] ,  \notag \\
f_{4} &=&
\begin{array}{l}
-\frac{\chi ge^{-i\omega _{a}t}}{\Delta \omega _{1}}\left[ \frac{%
e^{-i(\Delta \omega _{1}-\Delta \omega _{2})t}-1}{\Delta \omega _{1}-\Delta
\omega _{2}}+\frac{e^{i\Delta \omega _{2}t}}{\Delta \omega _{2}}\right]  \\
-\frac{\chi ge^{-i\omega _{a}t}}{\Delta \omega _{2}}\left[ \frac{%
e^{-i(\Delta \omega _{1}-\Delta \omega _{2})t}-1}{\Delta \omega _{1}-\Delta
\omega _{2}}-\frac{e^{-i\Delta \omega _{1}t}}{\Delta \omega _{1}}\right] ,
\end{array}
\label{f} \\
f_{5} &=&\frac{g^{2}e^{-i\omega _{a}t}}{\Delta \omega _{1}^{2}}\left[
e^{-i\Delta \omega _{1}t}-1\right] +\frac{ig^{2}te^{-i\omega _{a}t}}{\Delta
\omega _{1}},  \notag \\
f_{6} &=&f_{5},  \notag \\
f_{7} &=&\frac{\chi ^{2}e^{-i\omega _{a}t}}{\Delta \omega _{2}^{2}}\left[
e^{i\Delta \omega _{2}t}-1\right] -\frac{i\chi ^{2}te^{-i\omega _{a}t}}{%
\Delta \omega _{2}},  \notag \\
f_{8} &=&-f_{7}.  \notag
\end{eqnarray}

\begin{eqnarray}
g_{1} &=&\exp (-i\omega _{b}t),  \notag \\
g_{2} &=&-\frac{ge^{-i\omega _{b}t}}{\Delta \omega _{1}}\left[ e^{i\Delta
\omega _{1}t}-1\right] ,  \notag \\
g_{3} &=&
\begin{array}{l}
\frac{\chi ge^{-i\omega _{b}t}}{\Delta \omega _{2}(\Delta \omega _{1}-\Delta
\omega _{2})}\left[ e^{i(\Delta \omega _{1}-\Delta \omega _{2})t}-1\right]
\\
-\frac{\chi ge^{-i\omega _{b}t}}{\Delta \omega _{2}\Delta \omega _{1}}\left[
e^{i\Delta \omega _{1}t}-1\right] ,
\end{array}
\label{g} \\
g_{4} &=&
\begin{array}{l}
\frac{\chi ge^{-i\omega _{b}t}}{\Delta \omega _{2}(\Delta \omega _{1}+\Delta
\omega _{2})}\left[ e^{i(\Delta \omega _{1}+\Delta \omega _{2})t}-1\right]
\\
-\frac{\chi ge^{-i\omega _{b}t}}{\Delta \omega _{2}\Delta \omega _{1}}\left[
e^{i\Delta \omega _{1}t}-1\right] ,
\end{array}
\notag \\
g_{5} &=&\frac{g^{2}e^{-i\omega _{b}t}}{\Delta \omega _{1}^{2}}\left[
e^{i\Delta \omega _{1}t}-1\right] -\frac{ig^{2}te^{-i\omega _{b}t}}{\Delta
\omega _{1}},  \notag \\
g_{6} &=&-g_{5}.  \notag
\end{eqnarray}

\begin{eqnarray}
h_{1} &=&\exp (-i\omega _{c}t)  \notag \\
h_{2} &=&-\frac{ge^{-i\omega _{c}t}}{\Delta \omega _{1}}\left[ e^{i\Delta
\omega _{1}t}-1\right]   \notag \\
h_{3} &=&-\frac{\chi e^{-i\omega _{c}t}}{\Delta \omega _{2}}\left[
e^{i\Delta \omega _{2}t}-1\right]   \notag \\
h_{4} &=&
\begin{array}{l}
\frac{\chi ge^{-i\omega _{c}t}}{\Delta \omega _{2}}\left[ \frac{e^{i(\Delta
\omega _{1}+\Delta \omega _{2})t}-1}{\Delta \omega _{1}+\Delta \omega _{2}}-%
\frac{e^{i\Delta \omega _{1}t}}{\Delta \omega _{1}}\right]  \\
-\frac{\chi ge^{-i\omega _{c}t}}{\Delta \omega _{1}}\left[ \frac{e^{i(\Delta
\omega _{1}+\Delta \omega _{2})t}-1}{\Delta \omega _{1}+\Delta \omega _{2}}-%
\frac{e^{i\Delta \omega _{2}t}}{\Delta \omega _{2}}\right]
\end{array}
\label{h} \\
h_{5} &=&-\frac{g^{2}e^{-i\omega _{c}t}}{\Delta \omega _{1}^{2}}\left[
e^{i\Delta \omega _{1}t}-1\right] +\frac{ig^{2}te^{-i\omega _{c}t}}{\Delta
\omega _{1}}  \notag \\
h_{6} &=&-h_{5}  \notag \\
h_{7} &=&-\frac{\chi ^{2}e^{-i\omega _{c}t}}{\Delta \omega _{2}^{2}}\left[
e^{i\Delta \omega _{2}t}-1\right] +\frac{i\chi ^{2}te^{-i\omega _{c}t}}{%
\Delta \omega _{2}}  \notag \\
h_{8} &=&\frac{\chi ^{2}e^{-i\omega _{c}t}}{\Delta \omega _{2}^{2}}\left[
e^{i\Delta \omega _{2}t}-1\right] -\frac{i\chi ^{2}te^{-i\omega _{c}t}}{%
\Delta \omega _{2}}  \notag
\end{eqnarray}

\begin{eqnarray}
l_{1} &=&\exp (-i\omega _{d}t)  \notag \\
l_{2} &=&\frac{\chi e^{-i\omega _{d}t}}{\Delta \omega _{2}}\left[
e^{-i\Delta \omega _{2}t}-1\right]   \notag \\
l_{3} &=&
\begin{array}{l}
\frac{\chi ge^{-i\omega _{d}t}}{\Delta \omega _{1}(\Delta \omega _{1}-\Delta
\omega _{2})}\left[ e^{i(\Delta \omega _{1}-\Delta \omega _{2})t}-1\right]
\\
+\frac{\chi ge^{-i\omega _{d}t}}{\Delta \omega _{2}\Delta \omega _{1}}\left[
e^{-i\Delta \omega _{2}t}-1\right]
\end{array}
\label{l} \\
l_{4} &=&
\begin{array}{l}
\frac{\chi ge^{-i\omega _{d}t}}{\Delta \omega _{1}(\Delta \omega _{1}+\Delta
\omega _{2})}\left[ e^{-i(\Delta \omega _{1}+\Delta \omega _{2})t}-1\right]
\\
-\frac{\chi ge^{-i\omega _{d}t}}{\Delta \omega _{2}\Delta \omega _{1}}\left[
e^{-i\Delta \omega _{2}t}-1\right] \,
\end{array}
\notag \\
l_{5} &=&\frac{i\chi ^{2}te^{-i\omega _{d}t}}{\Delta \omega _{2}}+\frac{\chi
^{2}e^{-i\omega _{d}t}}{\Delta \omega _{2}^{2}}\left[ e^{-i\Delta \omega
_{2}t}-1\right]   \notag \\
l_{6} &=&l_{5}  \notag
\end{eqnarray}


\begin{thebibliography}{99}
\bibitem{foot2}  Same values of $\chi,$ $g$ and $|\alpha_{i}|$ are used in
the entire paper (unless otherwise specified). For spontaneous and partially
spontaneous Raman proceses these values of $|\alpha_{i}|$ are used used for
non-zero $|\alpha_{i}|$'s.
\end{thebibliography}
\end{document}